\documentclass[aip,jcp,reprint]{revtex4-1}

\usepackage{graphicx}
\usepackage{amsmath}
\usepackage{dcolumn}%
\usepackage[colorlinks=true,linkcolor=blue,citecolor=blue,urlcolor=blue]{hyperref}
\makeatletter
\@ifpackageloaded{hyperref}
  {\newcommand{\hrefdoi}[1]{, \href{http://dx.doi.org/#1}{DOI: #1}}}
  {\newcommand{\hrefdoi}[1]{}}
\@ifpackageloaded{hyperref}
  {\newcommand{\hrefarxiv}[2]{, \href{http://arxiv.org/abs/#1}{arXiv:#1 [#2]}}}
  {\newcommand{\hrefarxiv}[2]{}}
\makeatother

\usepackage{url}
\usepackage{upgreek}

\begin{document}

\title{Fingerprints of energy dissipation for exothermic surface chemical reactions: O$_2$ on Pd(100)}

\author{Vanessa J. Bukas}
\email{vanessa.bukas@ch.tum.de}
\affiliation{Chair for Theoretical Chemistry and Catalysis Research Center, Technische Universit{\"a}t M{\"u}nchen,\\ Lichtenbergstr. 4, D-85747 Garching, Germany}

\author{Shubhrajyoti Mitra}
\affiliation{Chair for Theoretical Chemistry and Catalysis Research Center, Technische Universit{\"a}t M{\"u}nchen,\\ Lichtenbergstr. 4, D-85747 Garching, Germany}

\author{J{\"o}rg Meyer}
\affiliation{Leiden Institute of Chemistry, Gorlaeus Laboratories, Leiden University, P.O. Box 9502, 2300 RA Leiden, The Netherlands}

\author{Karsten Reuter}
\affiliation{Chair for Theoretical Chemistry and Catalysis Research Center, Technische Universit{\"a}t M{\"u}nchen,\\ Lichtenbergstr. 4, D-85747 Garching, Germany}
\affiliation{SUNCAT Center for Interface Science and Catalysis, SLAC National Accelerator Laboratory \& Stanford University (USA),
443 Via Ortega, Stanford, CA 94035-4300, U.S.A.}

\date{\today}

\begin{abstract}
We present first-principles calculations of the sticking coefficient of O$_2$ at Pd(100) to assess the effect of phononic energy dissipation on this kinetic parameter. For this we augment dynamical simulations on six-dimensional potential energy surfaces (PESs) representing the molecular degrees of freedom with various effective accounts of surface mobility. In comparison to the prevalent frozen-surface approach energy dissipation is found to qualitatively affect the calculated sticking curves. At the level of a generalized Langevin oscillator (GLO) model we achieve good agreement with experimental data. The agreement is similarly reached for PESs based on two different semi-local density-functional theory functionals. This robustness of the simulated sticking curve does not extend to the underlying adsorption mechanism, which is predominantly directly dissociative for one functional or molecularly trapped for the other. Completely different adsorption mechanisms therewith lead to rather similar sticking curves that agree equally well with the experimental data. This highlights the danger of the prevalent practice to extract corresponding mechanistic details from simple fingerprints of measured sticking data for such exothermic surface reactions.
\end{abstract}

\pacs{34.35.+a,68.49.Df,82.20.Kh}

\maketitle
\section{Introduction}\label{sec:1}

Accurate knowledge of the sticking coefficient of molecules at transition metal surfaces is generally valuable in view of the relevance of this kinetic parameter in a wide range of applications such as heterogeneous catalysis. On more reactive surfaces (often desirable) dissociation can already happen alongside the adsorption process. Sticking data may then further provide information on the underlying adsorption mechanism. On the other hand, a sticking coefficient is an averaged kinetic quantity, comprising contributions from initially varying translational and rotational molecular impingement. To what extent sticking data really reflects mechanistic details is thus an open question \cite{King1984}. From a modeling perspective, there is vice versa the question of how well these underlying details need to be described to arrive at reliable sticking coefficients that can in turn e.g. be employed in microkinetic models \cite{Reuter2012}.

Aiming to answer these questions, intense fundamental research has been conducted to obtain an atomic-scale understanding of the reaction pathways and dynamics governing the adsorption process. Supersonic molecular beam techniques at low-index single crystal surfaces have made an unequivocal contribution in this respect, providing sticking coefficient data for a wide range of well-defined initial conditions of incidence energy and angle, as well as resolved rotational or vibrational state \cite{Vattuone2013}. In terms of dissociative adsorption corresponding data has been traditionally analyzed to distinguish regimes of direct or indirect adsorption \cite{Mullins1998}. The former essentially suggests immediate dissociation upon impact with the surface. The latter alludes to the notion of equilibration with the surface in a temporary intermediate (molecular) state which serves as a precursor to dissociation. Detailed studies of prototypical diatomics have especially served to establish general trends for sticking curves, i.e. certain signatures are seen to indicate one or the other adsorption mechanism \cite{Mullins1997}. A dependence of sticking data on substrate temperature is for instance believed to reflect precursor-mediated adsorption, considering the kinetic competition between dissociation and desorption from such a precursor state.

Such trends have met an overall intriguing consistency over a range of studied systems. Notwithstanding, one needs to recognize that a bulk of corresponding work has been done for (dissociative) H$_2$ adsorption, where a clearcut distinction might be facilitated by the relatively ``smooth'' molecule-surface interaction. Upon presence of $\pi$-orbital involving molecular bonds, the corresponding potential energy surface (PES) becomes significantly more intricate, which could prohibit such a direct extraction of mechanistic insight from measured sticking data. A characteristic example for this is the adsorption of N$_2$ on W(110), where measured sticking curves eluded a simple interpretation on the basis of the aforementioned experimental trends \cite{Mullins1997}. In fact, seemingly contradictory evidence from molecular beam measurements could only be reconciled through explicit dynamical simulations, which revealed a complex interplay between a direct and indirect adsorption mechanism depending on initial conditions \cite{Alducin2006}. The contribution of these processes was shown to arise from characteristics of the gas-surface interaction far from the surface and could only be captured within an accurate high-dimensional description of the PES involving all molecular degrees of freedom.

Similar complexities can be expected for the adsorption of O$_2$, a key process in oxidation catalysis. In fact, the typically large exothermocity of this process on transition metal surfaces adds yet another facet, namely the question of energy dissipation. Considering its influence e.g. on the aforementioned kinetic competition of dissociation or desorption from a precursor state, the question is how much this affects measured sticking curves and the established trends in terms of fingerprints. From a modeling perspective it again adds the question of how well the dissipative dynamics needs to be accounted for. The latter is particularly relevant, as a fully quantitative account of phononic energy dissipation in explicit \emph{ab initio} molecular dynamics (MD) simulations is still highly demanding, when aiming for sufficient statistical averaging and the full computation of energy-dependent sticking curves \cite{Gross2010,Meyer2014}. Highly appealing are therefore more effective treatments of surface mobility \cite{Hand1990,Adelman1976,Tully1980,Polanyi1985,Gross2015}. The potential sensitivity of O$_2$ sticking curves to the details of energy dissipation then makes a comparison to high-quality experimental data particularly valuable to gauge the accuracy of such effective approaches or vice versa the level of detail required to account for.

With this motivation we focus in the present work on the adsorption of O$_2$ on clean Pd(100). Recent molecular beam experiments found the initial sticking probability $S_0(E_{\rm i}, T_{\rm s})$ for this system to be independent of substrate temperature $T_{\rm s}$ and only weakly increasing with incident kinetic energy $E_{\rm i}$ \cite{Juurlink2015}. In terms of the classical trends, this suggested an interpretation in form of a predominantly direct dissociation mechanism with, at most, some reaction paths that include a modest activation barrier. At low $E_{\rm i}$ and $T_{\rm s}$, however, the contribution from a partly equilibrated molecular-precursor was conjectured in order to rationalize the independence on surface coverage and deviation from normal energy scaling under those conditions \cite{Juurlink2015}. We scrutinize this interpretation through dynamical simulations on a first-principles six-dimensional (6D) PES that accounts for all molecular degrees of freedom and which we suitably augment with effective treatments of surface mobility. Accounting in some respects for the latter is found to substantially change the calculated sticking curve $S_0(E_{\rm i}, T_{\rm s})$, i.e. the latter is indeed sensitive to energy dissipation. Intriguingly this holds for both direct and indirect adsorption mechanisms, either of which we obtain as dominant when basing the simulations on PESs obtained with two different density-functional theory (DFT) functionals. The uncertainties introduced by current semi-local DFT functionals thus prohibit a clear identification of the dominant adsorption mechanism. They also do not allow to fully disentangle whether the approximate treatment of substrate mobility or the deficiencies in the underlying DFT energetics are the primary reason for remaining small differences to the experimental data. All these intricacies nevertheless point already at this stage at the limitations of trying to directly deduce insight on the character of the adsorption process from measured sticking curves alone.

\section{Methods}\label{sec:2}

The O$_2$-Pd(100) interaction energetics are obtained by spin-polarized DFT calculations, using either the exchange correlation functional due to Perdew, Burke and Ernzerhof (PBE) \cite{PBE1996,PBE1997} or due to Hammer, Hansen and N{\o}rskov (RPBE) \cite{RPBE1999} to approximately assess the uncertainties introduced by prevalent semi-local functionals (see below). Electronic states are described with a plane wave basis set using a cut-off energy of 400\,eV as implemented in the {\tt CASTEP} code\cite{CASTEP}, together with ultrasoft pseudopotentials (USPPs) \cite{Vanderbilt1990} as bundled in the Materials Studio 6.0 database. These USPPs have been obtained with Vanderbilt's original generator \cite{Vanderbilt1990} using the PBE functional. By comparing to a few converged all-electron calculations based on the FHI-aims code \cite{Blum2009}, we have verified that the pseudopotential-induced error\cite{Fuchs1998,Kiejna2006} in the O$_2$-Pd(100) interaction energies is not larger than 100 meV \cite{Meyer2012}. Within a periodic supercell model, the surface is represented by five-layer slabs which are separated by a vacuum distance of 15\,{\AA} and which are (3$\times$3) multiples of the primitive surface unit cell of Pd(100). Calculations are performed using a (4$\times$4$\times$1) Monkhorst-Pack grid \cite{Monkhorst1976} for k-point sampling.

More than 6,000 DFT energies are calculated for various high- and low-symmetry configurations of the oxygen molecule above the frozen Pd(100) surface. These provide the basis for constructing a continuous representation of the adiabatic PES within all six molecular degrees of freedom ($V_{\mathrm{6D}}$) for each of the two DFT functionals based on symmetry-adapted neural networks as pioneered by Behler and Reuter~\cite{Reuter2007} for fcc(111) surfaces. Details about the present implementation can be found in Ref.~\cite{Meyer2012,Goikoetxea2012,Bukas2013}, including an adaption for fcc(100) surfaces in particular. We note that the resulting DFT-PBE PES has already been employed previously in Ref.~\onlinecite{Meyer2011,Meyer2012,Meyer2014}. Exactly the same recipe is followed here to obtain the continuous PES within the RPBE functional. See supplemental material for more details about the quality of both NN PES representations. Minima and barrier searches on these PESs have been conducted with a stochastic sampling method described in earlier work \cite{Bukas2013} and with the Nudged Elastic Band (NEB) method as implemented within the {\tt Atomic Simulation Environment} (ASE) \cite{Bahn2002}, respectively. 

Classical MD simulations are performed on the resulting numerically efficient continuous PES representations. The effect of including the initial zero point energy of the O$_2$ molecule within a quasi-classical treatment was additionally investigated but found to have a negligible effect on the resulting dynamics. Focusing only on the molecular degrees of freedom, such simulations do not allow for phononic energy dissipation and will henceforth be denoted with frozen surface (FS). A first account of surface mobility can be incorporated on the level of the 3D surface oscillator (SO) model \cite{Hand1990}. Here the surface is mimicked by an oscillator which is assigned the mass ($m_{\mathrm{SO}}$) of a single Pd atom and is permitted to move as a whole in all three directions within a harmonic potential. The associated $(3\times3)$ frequency matrix $\hat{\omega}_{SO}$ is assumed to be diagonal with values corresponding to a well localized surface mode of the Pd surface \cite{Meyer2012,Meyer2014}:
$\hbar\omega_{{\mathrm{SO}}_{\mathrm{xx}}} = \hbar\omega_{{\mathrm{SO}}_{\mathrm{yy}}} = 16 \, \mathrm{meV}$ and $\hbar\omega_{{\mathrm{SO}}_{\mathrm{zz}}} = 11 \, \mathrm{meV}$. The O$_2$-phonon coupling is then described by a 3D-space rigid shift $\mathbf{R}_{\mathrm{SO}}=(X_{\mathrm{SO}}, Y_{\mathrm{SO}}, Z_{\mathrm{SO}})$ of $V_{\mathrm{6D}}$, and the MD equations of motion in this approximation are given by
\begin{subequations}	\label{eq:SO}
\begin{align}
\frac{\partial^2 \mathbf{R}_{\mathrm{A,B}}}{\partial t^2} &= - \frac{1}{m_{\mathrm{A,B}}} \nabla_{\mathbf{R}_{\mathrm{A,B}}} V_{\mathrm{6D}} (\mathbf{R}_{\mathrm{A}}-\mathbf{R}_{\mathrm{SO}};\mathbf{R}_{\mathrm{B}}-\mathbf{R}_{\mathrm{SO}})  \\
\frac{\partial^2 \mathbf{R}_{\mathrm{SO}}}{\partial t^2} &= - \frac{1}{m_{\mathrm{SO}}} \nabla_{\mathbf{R}_{\mathrm{SO}}} V_{\mathrm{6D}} (\mathbf{R}_{\mathrm{A}}-\mathbf{R}_{\mathrm{SO}};\mathbf{R}_{\mathrm{B}}-\mathbf{R}_{\mathrm{SO}}) \nonumber \\
& \quad - {\hat{\omega}^2}_{\mathrm{SO}} \cdot \mathbf{R}_{\mathrm{SO}} \quad ,
\end{align}
\end{subequations}
where $\mathrm{m}_{\mathrm{A/B}}$ and $\mathbf{R}_{\mathrm{A/B}}$ are the masses and Cartesian coordinates of the two individual oxygen atoms A and B.

The effect of a bulk thermal bath is approximately included within the generalized Langevin oscillator (GLO) approach \cite{Adelman1976,Tully1980,Polanyi1985}. Here, the SO is coupled to a further 3D so-called ghost oscillator of equal mass, $m_{\rm GLO} = m_{\rm SO}$, and frequency matrix $\hat{\omega}_{\mathrm{GLO}} = \hat{\omega}_{\mathrm{SO}}$. The same frequencies are also used to describe the SO-GLO
coupling through a $(3 \times 3)$ coupling matrix $\hat{\Lambda}_{\rm SO-GLO} = \hat{\omega}_{\rm SO}$. The ghost oscillator is subject to frictional and random forces in order to account for energy dissipation and thermal fluctuations, respectively. As originally proposed by Adelman and Doll \cite{Adelman1976}, the former are described through an isotropic and diagonal damping matrix $\hat{\gamma}_{\mathrm{GLO}} =\gamma_{\rm GLO} \hat{\bf 1} = \pi \omega_{\mathrm{D}}/6 \, \hat{\bf 1}$, where $\omega_{\mathrm{D}}$ is the Pd bulk Debye frequency \cite{Rayne1964,Meyer2012}. Finally, the random force is a Gaussian white noise source $W$ with a variance of $(2k_{\mathrm{B}}T_{\mathrm{s}}\gamma_{\rm GLO} / m_{\mathrm{GLO}} \Delta t)^{1/2}$, where $k_{\mathrm{B}}$ is the Boltzmann constant and $\Delta t$ is the MD time integration step. The resulting equations of motion within the GLO model are thereby as follows:
\begin{subequations}	\label{eq:GLO}
\begin{align}
\frac{\partial^2 \mathbf{R}_{\mathrm{A,B}}}{\partial t^2} &= - \frac{1}{m_{\mathrm{A,B}}} \nabla_{\mathbf{R}_{\mathrm{A,B}}} V_{\mathrm{6D}} (\mathbf{R}_{\mathrm{A}}-\mathbf{R}_{\mathrm{SO}};\mathbf{R}_{\mathrm{B}}-\mathbf{R}_{\mathrm{SO}})  \\
\frac{\partial^2 \mathbf{R}_{\mathrm{SO}}}{\partial t^2} &= - \frac{1}{m_{\mathrm{SO}}} \nabla_{\mathbf{R}_{\mathrm{SO}}} V_{\mathrm{6D}} (\mathbf{R}_{\mathrm{A}}-\mathbf{R}_{\mathrm{SO}};\mathbf{R}_{\mathrm{B}}-\mathbf{R}_{\mathrm{SO}}) \nonumber \\
& \quad - {\hat{\omega}^2}_{\mathrm{SO}} \cdot \mathbf{R}_{\mathrm{SO}} + \hat{\Lambda}_{\mathrm{SO-GLO}} \cdot \mathbf{R}_{\mathrm{GLO}}   \\
\frac{\partial^2 \mathbf{R}_{\mathrm{GLO}}}{\partial t^2} &= - {\hat{\omega}^2}_{\mathrm{GLO}} \cdot \mathbf{R}_{\mathrm{GLO}} 
+ \hat{\Lambda}_{\mathrm{SO-GLO}} \cdot \mathbf{R}_{\mathrm{SO}} \nonumber \\
& \quad - \hat{\mathrm{\gamma}}_{\mathrm{GLO}} \frac{\partial \mathbf{R}_{\mathrm{GLO}}}{\partial t} + W(\Delta t)   \quad .
\end{align}
\end{subequations}
We note that this GLO implementation and specific choice of parameters follows that of earlier works regarding H$_2$ adsorbing on or scattered from the Pd(111) and Pd(110) surfaces \cite{Busnengo2001, Busnengo2004, Busnengo2005}. We nevertheless systematically tested the dependence on the specific parameter values by varying the oscillation frequencies entering $\hat{\omega}_{\mathrm{SO}}$, $\hat{\omega}_{\mathrm{GLO}}$, and $\hat{\Lambda}_{\mathrm{SO-GLO}}$ by one order of magnitude, by varying the damping coefficient $\gamma_{\rm GLO}$ by two orders of magnitude, as well as by doubling the mass $m_{\mathrm{SO}}$. This had little effect on the simulation results as will be further specified below.

The initial sticking coefficient at normal O$_2$ incidence was determined from  classical MD trajectories with the O$_2$ molecule initially with its center of mass at a distance $Z = 9$\,{\AA} from the surface, where the PES value for oxygen at its equilibrium bond length $d_{\mathrm{eq}} = 1.24$\,{\AA} is zero in both the DFT-PBE and DFT-RPBE PES representations. The initial molecular orientation and lateral center of mass position were sampled using a conventional Monte Carlo procedure. All statistical quantities are obtained by averaging over 5,000 trajectories for each value of incidence energy $E_{\mathrm{i}}$ and substrate temperature $T_{\mathrm{s}}$. Individual trajectories were integrated up to $10\,{\mathrm{ps}}$ in order to reach the following classification in terms of molecular adsorption, dissociative adsorption or reflection: A trajectory was classified as dissociative whenever the O$_2$ internuclear distance $d$ reached twice its equilibrium value ($d \geq 2d_{\mathrm{eq}}$) and is further increasing at this time ($\dot{d} > 0$), while reflection was concluded when the molecular center reached its initial starting distance above the surface with a positive $Z$-velocity. All trajectories where neither dissociation nor reflection occurred up to the $10\,{\mathrm{ps}}$ integration time were classified as trapped.

\section{Results and Discussion}\label{sec:3}
\subsection{Experimental sticking versus dynamics within the frozen surface approximation}\label{sec:3a}

\begin{figure}
\includegraphics{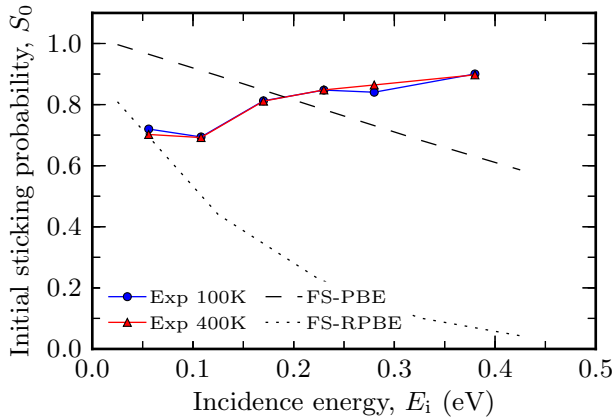}
\caption{(Color online) Initial sticking probability $S_0(E_{\rm i},T_{\rm s})$ of O$_2$ on Pd(100) as a function of incidence energy $E_{\rm i}$ and at normal angle of incidence. Experimental data for two substrate temperatures, $T_{\rm s} = 100$\,K (solid blue line) and 400\,K (solid red line), is reproduced from Ref.~\onlinecite{Juurlink2015}. Theoretical sticking probabilities are calculated in the temperature-independent frozen-surface (FS) approximation, using either DFT-PBE (dashed black line) or DFT-PRBE (dotted black line) energetics.}
\label{fig:fig1}
\end{figure}

Figure~\ref{fig:fig1} shows the initial sticking probability $S_0(E_{\rm i},T_{\rm s})$ at normal incidence as recently measured by molecular beam experiments \cite{Juurlink2015}. Sticking is generally high ($\geq70\%$) within the entire range of investigated incidence energies and largely identical at the two substrate temperatures investigated, 100\,K and 400\,K. The overall weak increase with incidence energy and independence of substrate temperature was interpreted to reflect predominantly direct dissociative adsorption \cite{Juurlink2015}. From additional data taken at finite coverages and oblique angles of incidence, some contribution from a partly equilibrated molecular-precursor was nevertheless speculated at low $(E_{\rm i},T_{\rm s})$.

\begin{figure}
\includegraphics{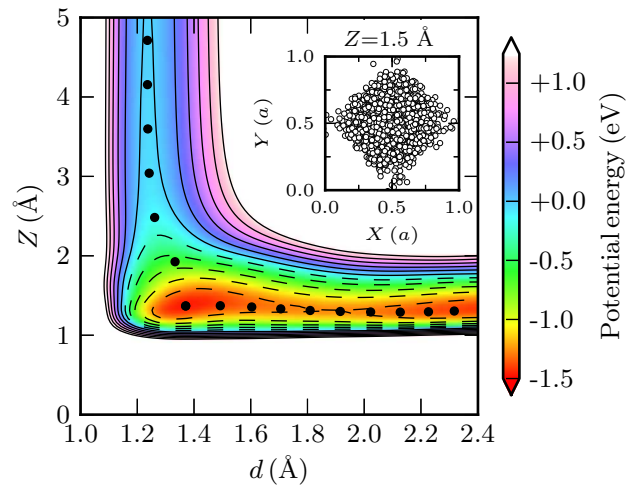}
\caption{(Color online) Contour plot of a 2D cut through the 6D O$_2$/Pd(100) DFT-PBE PES depicting the existence of a pronounced molecular chemisorption well. Shown is the energy profile as a function of the vertical distance of the O$_2$ center of mass $Z$ and the O$_2$ bond length $d$ for a O$_2$ molecule in a side-on configuration with the molecular axis oriented along the [001] direction. Additionally shown by black markers is the minimum energy path towards dissociation as obtained from a NEB calculation. Black contour lines indicate $250\,\mathrm{meV}$ energy increments. Inset: Lateral O$_2$ center of mass positions $(X,Y)$ over the surface unit-cell for the ultimately dissociating trajectories as molecules with $E_{\rm i} = 25$\,meV first reach a surface distance of $Z=1.5\,${\AA} (see text).}
\label{fig:fig2}
\end{figure}

Even without explicit dynamical simulations the overall high reactivity can already be gleaned directly from the calculated O$_2$-Pd(100) interaction potential. From a global search on the continuous 6D DFT-PBE PES representation \cite{Bukas2013}, we identify as most stable molecular adsorption state a configuration in which the O$_2$ molecule centers side-on above the Pd(100) hollow site with its molecular axis oriented along the [001] direction, i.e. in the direction of the neighboring hollow sites. This molecular well can be reached along a barrierless entrance channel as visible in the 2D (elbow) cut through the PES shown in Fig.~\ref{fig:fig2}. The calculated minimum energy path to dissociation also included in Fig.~\ref{fig:fig2} exhibits a dissociation barrier out of this well of $E_{\rm diss} =200$\,meV. This is a small value compared to the actual depth of the molecular well of $E_{\rm b} = -1.45$\,eV with respect to the gas phase. Even if a fraction of this energy gained upon adsorption is quickly transferred to internal molecular degrees of freedom, such a small barrier should thus be readily surmounted. Based on highly consistent DFT-PBE energetics ($E_{\rm b} = -1.52$\,eV, $E_{\rm diss} = 120$\,meV), this was the conclusion taken by Liu and Evans, who correspondingly classified the dissociative adsorption process as essentially non-activated \cite{Evans2014}. The overall high reactivity measured in the experiment seems to support this assignment, as well as the general notion to assume an approximately unity sticking coefficient in coarse-grained microkinetic simulations whenever static total energy calculations identify a barrierless entrance channel to adsorption \cite{Reuter2012, Scheffler2006, Scheffler2007, Scheffler2008}.

Aiming to scrutinize such mechanistic assignments made on the basis of the static PES alone we proceed with the classical trajectory calculations. Evaluating the fraction of reflected and adsorbing trajectories yields the theoretical sticking coefficient at a given incidence energy. Within the frozen-surface approximation, i.e. restriction to the 6D PES, the resulting sticking curve $S_0(E_{\rm i})$ is necessarily independent of substrate temperature and is compared to the experimental sticking curves in Fig.~\ref{fig:fig1}. Confirming the expectations from the attractive DFT-PBE PES the theoretical sticking is rather high at low $E_{\rm i}$ and therewith in the same ballpark as the experimental data. For these low incidence energies there is thus no qualitative disagreement with experiment at the level of semi-local DFT as e.g. in case of the enigmatic O$_2$ at Al(111) system \cite{Behler2005,Libisch2012}. This is not too surprising considering that a major break-down of the electronic adiabaticity of the adsorption process had been invoked as one possible reason behind the O$_2$/Al(111) discrepancy \cite{Behler2005,Carbogno2013}. This unlikely applies to Pd(100), which exhibits a very high electronic density-of-states at the Fermi level and for which only small energy losses to electron-hole pairs during O$_2$ adsorption have been calculated \cite{Meyer2011}.

In terms of overall (rough) magnitude the favorable agreement of experimental and theoretical sticking extends over the entire range of incidence energies shown in Fig. 1. Nevertheless, there is a disturbing difference in trend with $E_{\rm i}$. In contrast to the measured slight rise of sticking with $E_{\rm i}$, the calculated FS sticking curve shows a continuous decline. This decline is in fact even more pronounced when calculating the sticking curve on the basis of the DFT-RPBE PES, cf. Fig.~\ref{fig:fig1}. Repeating the above static analysis also for this PES we find its topology to be qualitatively very similar to that of the DFT-PBE PES. Important quantitative differences are therefore nicely summarized in terms of the two dynamically relevant quantities, $E_{\rm b}$ and $E_{\rm diss}$. In line with the general construction idea of the RPBE functional \cite{RPBE1999}, the molecular well is much more shallow in the DFT-RPBE PES ($E_{\rm b} = -0.85$\,eV), while reflecting the concomitant weaker bond activation the dissociation barrier is with $E_{\rm diss} = 400$\,meV about twice as high as at the DFT-PBE level. Even though the ratio of these two quantities is thus much less favorable, on purely energetic grounds dissociative adsorption would nevertheless still be classified as non-activated even at the DFT-RPBE level, i.e. the molecules gain much more energy upon adsorption than is needed to overcome the dissociation barrier.

The decline of the sticking curve with $E_{\rm i}$ obtained with both functionals is thus a purely dynamical effect. Aiming to extract its origin we analyze the trajectories in more detail. First of all, this trajectory analysis largely confirms the dynamical relevance of the minimum energy path depicted in Fig.~\ref{fig:fig2}. Independent of the incidence energy and for both functionals (as well as all later surface mobility treatments), essentially all trajectories leading to adsorption show the molecules accumulating first around the hollow molecular chemisorption well. This preference for adsorption above hollow is exemplified in the inset of Fig.~\ref{fig:fig2} which analyzes the lateral O$_2$ center of mass positions over the surface unit-cell for the ultimately dissociating dynamical trajectories as the molecules first reach a surface distance of $Z=1.5\,${\AA}. The data shown corresponds to $E_{\rm i} = 25$\,meV and the PBE functional, with equivalent findings obtained at all other incidence energies and for DFT-RPBE.

\begin{figure}
\includegraphics{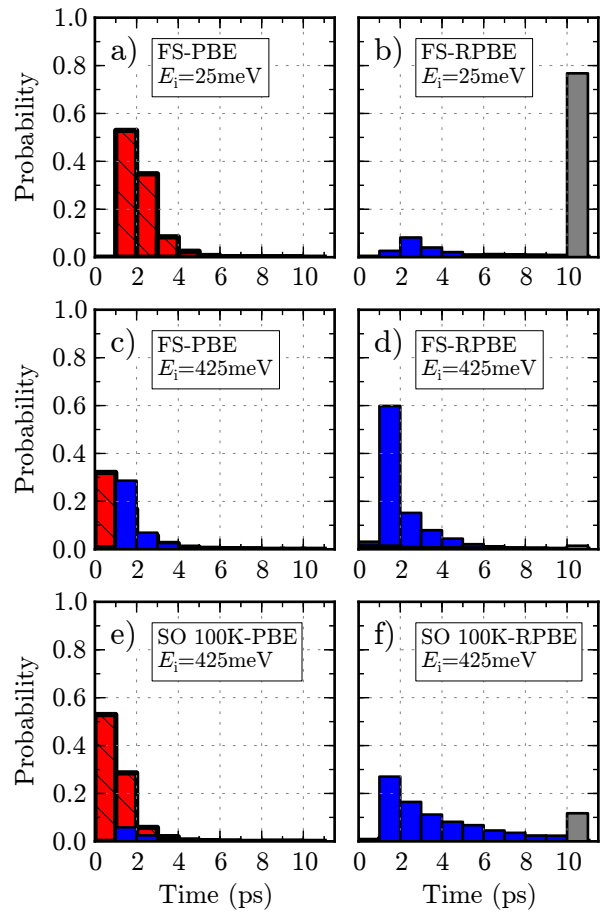}
\caption{(Color online) Time distribution characterizing events of reflection (blue), direct dissociation (shaded red) and molecular trapping (gray) as classified within the performed MD simulations. Trapping is assigned if neither dissociation nor reflection has occurred after 10\,ps simulation time, which is why the corresponding fraction is shown at this time bin. Left and right panels refer to trajectories calculated on the basis of the DFT-PBE and DFT-RPBE PES descriptions, respectively, while the employed level of theory and initial conditions ($E_{\rm i}$ and $T_{\rm s}$) are as reported in the corresponding panel labels. We note that summing up the red and gray bars yields the initial sticking probabilities $S_0$ of the respective functional at the particular incidence energies $E_{\rm i}$ and surface temperatures $T_{\rm s}$ shown in Figs~\ref{fig:fig1} and \ref{fig:fig4}.}
\label{fig:fig3}
\end{figure}

After arrival at the molecular chemisorption state, the fate of the molecules is almost instantaneously decided at low $E_{\rm i}$ at the DFT-PBE level. This is demonstrated by the time distribution shown in Fig. \ref{fig:fig3}a, which indicates after which time a trajectory has been classified as dissociated, reflected or trapped. Recall that trapping is assigned to all molecules that have neither dissociated nor reflected after 10\,ps simulation time, which is why the corresponding fraction is shown at this time bin in Fig. \ref{fig:fig3}. As apparent from Fig. \ref{fig:fig3}a, essentially all trajectories dissociate in a rather direct fashion, i.e. within several picoseconds and thus after a minimal number of surface rebounds (typically $<5$). This is completely different at the DFT-RPBE level, where almost all trajectories instead end up being trapped, cf. Fig. \ref{fig:fig3}b. The different reactivity of the two DFT PESs ($E_{\rm b}$ and $E_{\rm diss}$) thus does have a significant effect in terms of entirely changing the dominant adsorption mechanism from direct (DFT-PBE) to indirect (DFT-RPBE). At low incidence energies this does not show up in the sticking coefficient though, as already a small amount of energy transfer to internal vibrational, rotational or lateral translational degrees of freedom is sufficient to prevent immediate desorption, and since molecules correspondingly trapped in the surface potential are being counted as contributing towards sticking. We note that in the absence of any other dissipation mechanism within the FS approximation, such trapped molecules would in fact eventually be able to desorb. Notwithstanding, the time scale for a corresponding transfer of energy back into the perpendicular translational motion would be rather long. In reality other dissipation channels would have set in during this time, which is why it is reasonable to count these trajectories as contributing to the sticking coefficient. Also the rather arbitrarily set maximum integration time of 10\,ps should in this respect not matter, as we have verified by applying the trapping classification already after 8\,ps.

While the PES differences thus do not show up at low $E_{\rm i}$, they do increasingly at larger incidence energies. Corresponding fast molecules are less efficiently steered to direct dissociation. Neither is the energy transfer to internal degrees of freedom efficient enough anymore to quickly remove the large excess kinetic energy and trap them in the surface potential. As shown in Figs. \ref{fig:fig3}c and d for the largest $E_{\rm i} = 425$\,meV we thus find for both PES descriptions a large number of molecules that are reflected very quickly from the surface. Without efficient trapping, only the decreasing fraction of molecules that already starts from configurations favorable for a more or less direct dissociation mechanism can contribute to the decreasing sticking seen in Fig.~\ref{fig:fig1}. Such a mechanism is generally disfavored by a smaller acceleration into a more shallow adsorption well and a higher barrier to surmount, which is why the concomitant decline of $S_0(E_{\rm i})$ with $E_{\rm i}$ is also much more pronounced at the DFT-RPBE level.

From a bulk of work on O$_2$ adsorption on late transition metals \cite{Carbogno2013}, the PBE and RPBE functionals can be seen as popular representatives for opposite ends within the range of current gradient-corrected functionals, with the prior likely more on the overbinding side and the latter possibly slightly underbinding. Within the understanding of the just presented dynamical analysis concomitant uncertainties in the energetic description of the molecular chemisorption well ($E_{\rm b}$ and $E_{\rm diss}$) at the semi-local DFT level are nevertheless unlikely to cause the wrong trend of the sticking curve with $E_{\rm i}$ as compared to the experimental data. Rather than the overall attractiveness of the PES, the major reason for the decline of the calculated sticking curves lies in an insufficient ability to trap the molecules in the chemisorption well. This thus points more to a weakness of the FS approximation, in which such a trapping can only result from energy transfer into molecular rovibrational degrees of freedom. In reality, this can also come from inelastic collisions with the surface atoms, which suggests the necessity to include some degree of surface mobility into the modeling.

\subsection{Surface mobility on the level of SO and GLO models}\label{sec:3b}

\begin{figure}
\includegraphics{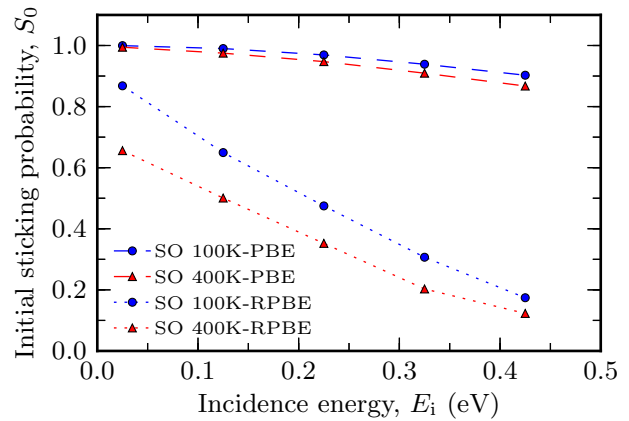}
\caption{(Color online) Comparison of the initial sticking probability $S_0(E_{\rm i},T_{\rm s})$ of O$_2$ on Pd(100) as calculated within the surface oscillator model (SO) for normal angle of incidence and varying incidence energy $E_{\rm i}$. Data are shown for substrate temperatures of $T_{\rm s}=100$\,K (blue circles) and 400\,K (red triangles), using either DFT-PBE (dashed lines) or DFT-PRBE (dotted lines) energetics.}
\label{fig:fig4}
\end{figure}

We introduce a first account of energy exchange with the lattice through the SO model and show the derived initial O$_2$/Pd(100) sticking probabilities for substrate temperatures of $T_{\rm s} = 100$\,K and 400\,K in Fig.~\ref{fig:fig4}. On the basis of the DFT-PBE energetics, and compared to the equivalent description within the FS approximation, we indeed observe the qualitative change expected from our preceding dynamical analysis, namely a considerable increase in high-energy sticking (up to as much as $25\%$ for $E_{\rm i} = 425$\,meV). This almost completely lifts the decline with $E_{\rm i}$ seen at the FS level. The resulting essentially constant $S_\mathrm{0}(E_{\rm i}, T_{\rm s})$ is furthermore only weakly dependent on $T_{\mathrm{s}}$ and thus in seemingly improved agreement with experiment.

The mechanistic analysis of the PBE-based adsorptive trajectories reveals the relevance of the same reaction pathway above hollow as found in the FS dynamics and depicted in Fig.~\ref{fig:fig2}. Additionally, the short reaction/reflection times analyzed in Fig.~\ref{fig:fig3}e once again suggest that the fate of impinging molecules is almost instantaneously determined and direct dissociation takes place essentially upon first impact with the surface. The differences in reactivity compared to the FS case do therefore not originate from the activation of different reaction pathways or adsorption mechanisms. We instead rationalize the increased SO sticking at high $E_{\rm i}$ with the improved efficiency of direct dissociation above hollow due to an enhanced steering of impinging fast molecules in the now ``elastic'' surface potential. The energy transfer to the surface possible within the SO model simply dampens the perpendicular translational motion of highly accelerated O$_2$ already upon first impact. This increases the probability that they will be effectively captured by the attractive potential and subsequently led to dissociation. This picture arises from the analysis of the significant amount of trajectories that start from the exact same initial conditions and yet are reflected within the FS but dissociated within the SO model. The corresponding energy profiles reveal the simultaneous onset of the O$_2$-Pd(100) interaction and the SO kinetic energy, and therewith underline the decisive role of the SO energy uptake in ultimately determining the trajectory outcome.

Quantifying this SO energy uptake in fact reveals that considerable amounts of energy are transferred back and forth between the O$_2$ molecule and the SO even on the short time scales until dissociation. At the moment when the trajectories fulfill the dissociation criterion an average of 350\,meV (considering the sum of kinetic and potential energy of the oscillator) is stored in the SO for $T_{\rm s} = 100$\,K and a low incidence energy of 25\,meV. For the fast molecules with $E_{\rm i} = 425$\,meV this value even increases up to 510\,meV. In the absence of any further dissipation channel, the SO is thus severely overheated. As a result, one may expect that the promotion of direct dissociation over ``proper'' molecular trapping is to some degree artificial and that the agreement with the experimental sticking curve on the DFT-PBE SO level is thereby merely fortuitous. 

This view is indeed supported by the results obtained with the DFT-RPBE PES, where, as already indicated, dissociation is generally less favored due to the limited accessibility of the corresponding transition state. Here, the partial damping on account of the surface mobility also succeeds in trapping the molecules, but only for a slightly extended period of time compared to the FS situation. This is visible from the extended reflection time distribution in Fig.~\ref{fig:fig3}f compared to Fig.~\ref{fig:fig3}d. The ongoing strong energy exchange between trapped molecules and hot SO thus promotes in this case rather a delayed reflection than (equally artificially) facilitating dissociation. The resulting sticking curve at the DFT-RPBE SO level shown in Fig.~\ref{fig:fig4} correspondingly still exhibits the strong decline with incidence energy.

\begin{figure}
\includegraphics{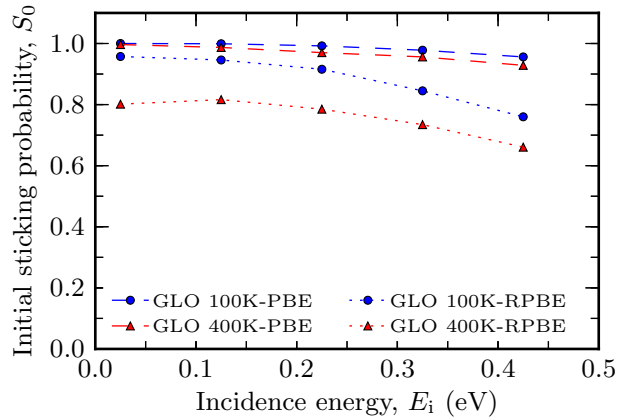}
\caption{(Color online) Comparison of the initial sticking probability $S_0(E_{\rm i},T_{\rm s})$ of O$_2$ on Pd(100) as calculated within the generalized Langevin oscillator model (GLO) for normal angle of incidence and varying incidence energy $E_{\rm i}$. Data are shown for substrate temperatures of $T_{\rm s}=100$\,K (blue circles) and 400\,K (red triangles), using either DFT-PBE (dashed lines) or DFT-PRBE (dotted lines) energetics.}
\label{fig:fig5}
\end{figure}

\begin{figure}
\includegraphics{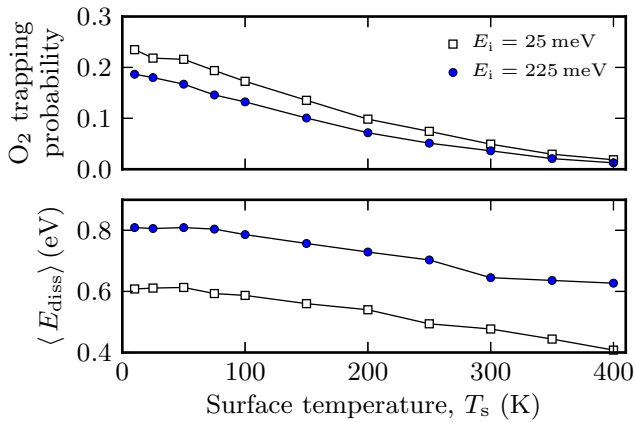}
\caption{(Color online) Top panel: Temperature dependence of the probability of molecular trapping within the GLO model and using the DFT-PBE energetics. Bottom panel: Average amount of energy dissipated into the GLO heat bath for the corresponding trapped trajectories. Data are shown for incident energies of $E_{\mathrm{i}} = 25$ (open squares) and $225\,\mathrm{meV}$ (closed circles).}
\label{fig:fig6}
\end{figure}

Suspecting the lack of further energy dissipation channels as a major limitation in the present application of the SO model we proceed with classical trajectory calculations that now include a coupling to a bulk thermal bath within the GLO approach. At first glance, application of the GLO model on the basis of the DFT-PBE PES has an almost inconsequential effect on sticking as compared to the results obtained within the SO approximation. Figure~\ref{fig:fig5} now shows an entirely constant unity sticking coefficient independent of both $E_{\rm i}$ and $T_{\rm s}$. Analysis of the underlying GLO trajectories nevertheless reveals again substantial changes that are fully consistent with the expectations from the SO analysis. Also at the DFT-PBE level there is now a contribution of molecular trapping to the total sticking, i.e. the GLO succeeds at least to some extent in removing the artificially enhanced direct dissociation seen at the SO level. The average amount of energy stored in the SO itself at the moment when trajectories fulfill the dissociation criterion is now reduced to 210\,meV (260\,meV) at $E_{\rm i} = 25$\,meV (425\,meV) and $T_{\rm s} = 100$\,K, as compared to the 350\,meV (510\,meV) found before within the SO model. Owing to the non-equilibrium nature of O$_2$ dissociation on Pd(100) \cite{Meyer2012}, this is still far away from the thermal equilibrium value of a surface atom $2 \cdot \tfrac{3}{2} k_{\rm B} T_{\rm s} \approx$ 25\,meV even within the limitations of the GLO model, but still suggests more efficient surface equilibration. The latter is also reflected in the dependence of the trapping contribution to the overall sticking on the $(E_{\rm i},T_{\rm s})$ conditions summarized in Fig.~\ref{fig:fig6}. As expected for a (largely) equilibrated molecular precursor this contribution depends only weakly on the initial incidence energy, but instead sensitively on the substrate temperature. As also shown in Fig.~\ref{fig:fig6}, the varying excess kinetic energy of molecules with different $E_{\rm i}$ is thus successfully drained into the heat bath, i.e. the average energy dissipated into the bath is for every substrate temperature precisely by 200\,meV higher for molecules impinging with $E_{\rm i} = 225$\,meV as compared to those impinging with $E_{\rm i} = 25$\,meV.

On the basis of the DFT-RPBE energetics molecular trapping remains the only adsorption mechanism also at the GLO level. Allowing for ``dissipated'' energy to leave the SO into the heat bath, however, now suppresses the artificial possibility that some fraction of this energy is returned to the adsorbate and thereby induces desorption. This removes the delayed reflection times observed within the equivalent SO results, cf. Fig.~\ref{fig:fig3}f, and yields a largely increased sticking coefficient at high incidence energies. The resulting DFT-RPBE GLO sticking curves shown in Fig.~\ref{fig:fig5} exhibit only a weak decline with $E_{\rm i}$. They also exhibit only a weak dependence on substrate temperature. This is rather intriguing in view of the predominance of the precursor mechanism, which is generally believed to fingerprint in form of a sensitive substrate temperature dependence. Overall, the DFT-RPBE GLO curve agrees therewith now rather well with the experimental data, at least similarly well as the DFT-PBE GLO sticking curve, cf. Fig.~\ref{fig:fig5}.

This is, in fact, a rather intriguing, if not disturbing result. At the GLO level both DFT-PBE and DFT-RPBE descriptions yield rather similar sticking curves, and this despite predicting completely different adsorption mechanisms: Predominantly direct dissociation with a small contribution of molecular trapping at low substrate temperatures within DFT-PBE versus completely indirect dissociation within DFT-RPBE. Compared to the experimental data both functional levels reach indeed the same achievements and exhibit the same shortcomings. They correctly predict a near-unity sticking over the range of studied incidence energies with if at all only a weak dependence on substrate temperature. Yet, both fail to reproduce the measured slight increase in sticking with $E_{\rm i}$, or more precisely they overestimate the experimental sticking at low $E_{\rm i}$. 

Dissipation seems to affect much more the high incidence energy part of the sticking curves. This view is also supported by the robustness of the obtained sticking results when varying the parameters entering the GLO model as described in Section II. All in all, these variations lead only to insignificant variations of $S_o(E_{\rm i}, T_{\rm s})$ within a few percent. In turn, the uncertainties in the semi-local DFT description of the actual chemisorption well ($E_{\rm b}$ and $E_{\rm diss}$) do critically affect the actual adsorption mechanism, yet again do not seem to propagate to the total sticking curve. In light of the dynamical analyses performed for the FS, SO and GLO models, we suggest that a slightly incorrect description of the entrance channel part of the PES can be one reason for the remaining small discrepancy -- within these effective treatments of dissipation. Presently unaccounted small activation barriers in this channel for some molecular configurations are likely to specifically affect the total sticking at low incidence energies and could therewith constitute  the ``missing ingredient'' to a fully quantitative agreement with experiment. 

Unless there is a problem in the experimental data, the obvious and systematic limitation of these models is an equally likely cause: Reducing the phononic fine structure of the Pd surface into computationally convenient augmentations of the frozen-surface within the O$_2$-Pd(100) interaction can not, by construction, account for the energy exchange with an entire layer of moving surface atoms. Only detailed future studies including all these degrees of freedom will allow to completely capture the influence of surface mobility on the calculated sticking curves.

\section{Summary and conclusions}\label{sec:4}

We presented a detailed calculation of the initial sticking coefficient of O$_2$ at Pd(100) based on classical trajectory calculations on first-principles 6D potential energy surfaces. The specific motivation was to elucidate the role of energy dissipation during the exothermic surface reaction by augmenting the 6D dynamics with various effective accounts of surface mobility, as well as to assess in how much classical trends in the sticking data reveal the underlying adsorption mechanism. To our knowledge for the first time, we observe that an account of energy dissipation leads to qualitative changes of the calculated sticking curve as compared to the prevalent frozen-surface approximation. Compared to the much more frequently studied adsorption of H$_2$ (also at Pd surfaces \cite{Busnengo2001, Busnengo2004, Busnengo2005}) this might not come as altogether surprising in view of the much smaller mass mismatch of oxygen with palladium. It is, however, remarkable with respect to the rather similar N$_2$ on W(110) system \cite{Alducin2006} ($\tfrac{m_{\mathrm{N}}}{m_{\mathrm{W}}} \approx \tfrac{14}{184}$). Intermediate in its mass ratio compared to H$_2$/Pd ($\tfrac{m_{\mathrm{H}}}{m_{\mathrm{Pd}}} \approx \tfrac{1}{106}$) and O$_2$/Pd ($\tfrac{m_{\mathrm{O}}}{{m_\mathrm{Pd}}} \approx \tfrac{16}{106}$), it exhibits PES characteristics intriguingly similar to that of O$_2$/Pd(100), i.e. non-activated paths to dissociation alongside a precursor molecular well. Yet, the N$_2$/W(110) sticking curve showed no change when applying effective treatments beyond the frozen-surface approach.

The hierarchical application of FS, SO and GLO models reveals how the calculated sticking curves sensitively respond to details of energy exchange with the substrate. Already some degree of energy exchange as provided by the SO model enhances the steering of highly accelerated impinging O$_2$, while ``proper'' molecular trapping necessitates an account of further bulk dissipation channels as in the GLO. The large amount of energy released in the exothermic reaction otherwise simply leads to an overheating of the SO degree of freedom.  Compared to the N$_2$/W(110) system, one important difference to note is the considerably deeper molecular well in the corresponding FS-PES description ($E_{\rm b} = -0.39$\,eV \cite{Bocan2008} and $-0.85$\,eV respectively within DFT-RPBE) which leads to a significantly increased acceleration of impinging molecules and requires more than twice as much energy to be dissipated upon reaching the molecular state. The latter may also hint towards a different response of the real phonons -- potentially mitigated by the different phononic properties of the bare surfaces alone. It might thus comprise a key discriminating characteristic, which could be further elucidated by more conceptually and computationally demanding studies including a more realistic phonon heat bath \cite{Meyer2014}.

For two different PES representations based on the DFT-PBE and DFT-RPBE functionals we correspondingly obtain at the GLO level a rather satisfying agreement with experimental sticking data \cite{Juurlink2015}. This in principle appealing robustness of the simulation results with respect to the uncertainties of semi-local DFT energetics does not extend to the level of the underlying adsorption mechanism though. The more attractive DFT-PBE energetics predicts a predominantly direct dissociation mechanism with some amount of molecular trapping at low substrate temperatures. The less attractive DFT-PRBE energetics instead predicts adsorption almost exclusively via the molecular precursor state.

Completely different adsorption mechanisms therewith lead to rather similar sticking curves that agree equally well with the experimental data. Independent of the small quantitative discrepancies that remain in either case with respect to experiment, this clearly demonstrates that an unambiguous deduction of the adsorption mechanism from the sticking data alone is not feasible for this system. Such fingerprinting may work for simpler adsorption systems with smoother potential energy surfaces. At the latest for reactions with high degree of exothermicity dedicated calculations explicitly accounting for high-dimensional potential energy surfaces including substrate mobility at best in the form of moving surface atoms are required to establish the mechanistic details. These details can still be very important not only in light of fundamental understanding but also for coarse-grained micro-kinetic models \cite{Liu2014,Liu2015}.

Completely different adsorption mechanisms therewith lead to rather similar sticking curves that agree equally well with the experimental data. Independent of the small quantitative discrepancies that remain in either case with respect to experiment, this clearly demonstrates that an unambiguous deduction of the adsorption mechanism from the initial sticking data alone is not feasible for this system. Such fingerprinting may work for simpler adsorption systems with smoother potential energy surfaces. At the latest for reactions with high degree of exothermicity dedicated calculations explicitly accounting for high-dimensional potential energy surfaces including substrate mobility at best in the form of moving surface atoms are required to establish the mechanistic details. These details can still be very important not only in light of fundamental understanding but also for coarse-grained micro-kinetic models \cite{Evans2014,Liu2015}.

\begin{acknowledgments}
We thank H. F. Busnengo for providing his SO and GLO implementations as well as many stimulating discussions. Funding through the Deutsche Forschungsgemeinschaft is acknowledged within project RE1509/19-1, as is generous access to CPU time through the Leibniz Rechenzentrum der Bayerischen Akademie der Wissenschaften (project pr85wa).
\end{acknowledgments}

\end{document}